\documentstyle[12pt,amstex,amssymb]{article}
\input{feynman}
\renewcommand{\textfraction}{0}
\renewcommand{\floatpagefraction}{1}
\let\sstl=\scriptscriptstyle
\def\Was{W\c as}
\def\Order#1{${\cal O}(#1$)}
\def\Ordpr#1{${\cal O}(#1)_{prag}$}
\def\bbe{\bar{\beta}}
\def\tbe{\tilde{\beta}}
\def\tal{\tilde{\alpha}}
\def\tom{\tilde{\omega}}
\def\half{ {1\over 2} }
\def\alf1{ {\alpha\over\pi} }

\begin{document}
 
\begin{titlepage}
 \begin{flushright}
 
 {\bf UTHEP-96-1101 }\\
 {\bf Nov.~~~ 1996}\\
\end{flushright}
\vspace{0.3cm}
 
\begin{center}
{\LARGE Exact ${\cal O}(\alpha)$
Gauge Invariant YFS Exponentiated\\ 
Monte Carlo for (Un)Stable 
$W^+W^-$ Production\\ 
At and Beyond LEP2 Energies$^{\dagger}$
}
\end{center}

\begin{center}
 {\bf S. Jadach}\\
   {\em Institute of Nuclear Physics,
        ul. Kawiory 26a, Krak\'ow, Poland}\\
   {\em CERN, Theory Division, CH-1211 Geneva 23, Switzerland,}\\
 {\bf W.  P\l{a}czek}$^{\star}$\\
   {\em Department of Physics and Astronomy,\\
   The University of Tennessee, Knoxville, Tennessee 37996-1200},\\
 {\bf M. Skrzypek}\\
   {\em Institute of Nuclear Physics,
        ul. Kawiory 26a, Krak\'ow, Poland}\\
 {\bf B.F.L. Ward}\\
   {\em Department of Physics and Astronomy,\\
   The University of Tennessee, Knoxville, Tennessee 37996-1200,\\
   SLAC, Stanford University, Stanford, California 94309,}\\
{\bf Z. W\c as}\\
   {\em Institute of Nuclear Physics,
        ul. Kawiory 26a, Krak\'ow, Poland}\\
   {\em CERN, Theory Division, CH-1211 Geneva 23, Switzerland}\\
\end{center}

\vspace{0.25cm}
\begin{center}
{\bf   Abstract}
\end{center}
 
We realize, by Monte Carlo event generator methods, the exact
${\cal O}(\alpha)$ YFS
exponentiated calculation of $e^+e^- \rightarrow
W^+ W^- (\rightarrow  f_1\bar f'_1 + \bar f_2 f'_2)$ at and beyond
LEP2 energies, where the left-handed parts of
$f_i$ and $f'_i$ are the respective upper and lower
components of an $SU_{2L}$ doublet, $i=1,2$.
Our calculation is gauge invariant from the standpoint of
its radiative effects and the respective YFS Monte Carlo
event generator YFSWW3, wherein both Standard Model and anomalous
triple gauge boson couplings are allowed, generates $n(\gamma)$
radiation both from the initial state and from the final $W^+ W^-$.
Sample Monte Carlo data are illustrated.

\vspace{0.2cm}
\begin{center}
{\it To be submitted to Physics Letters B}
\end{center}
 
\vspace{0.750cm}
\renewcommand{\baselinestretch}{0.1}
\footnoterule
\noindent
{\footnotesize
\begin{itemize}
\item[${\dagger}$]
Work partly supported by the Polish Government
grants KBN 2P30225206 and 2P03B17210 and
by the US Department of Energy Contracts  DE-FG05-91ER40627
and   DE-AC03-76ER00515.

\item[${}^{\star}$]
On leave of absence from Institute of Computer Science, Jagellonian
University, Krak\'ow, Poland.
\end{itemize}
}
 
\begin{flushleft} 
{\bf UTHEP-96-1101 }\\
{\bf Nov.~~~ 1996}\\
\end{flushleft}

\end{titlepage}

\section{Introduction}

The processes 
$e^+e^- \to W^+W^- +n(\gamma)\to 4 fermions+n(\gamma)$
at and beyond LEP2
energies are of considerable interest 
for the LEP2 and NLC physics programs. When calculated and measured
to sufficient precision, they provide windows to several important
avenues for
verification and tests of the $SU_{2L}\times U_1$ model of Glashow,
Salam and Weinberg~\cite{gsw} of the electroweak interaction.
In Ref.~\cite{LEP2YBWW} for example, the goal for the
theoretical precision of the understanding of these processes
was set at $0.5\%$ for the LEP2 physics program.  In the following
discussion, we will present calculations of these processes
which feature the most precise electroweak radiative corrections
which have been published to date, that of an exact ${\cal O}(\alpha)$
YFS~\cite{yfs} exponentiated Monte Carlo event generator
in which multiple photon radiatve effects from both initial and
$W^+W^-$ states are realized on an event-by-event basis as 
infrared singularities are canceled to all orders in $\alpha$.
We denote the respective event generator by YFSWW3. We will comment
below on its physical and technical precision relative to the
$0.5\%$ precision goal of Ref.~\cite{LEP2YBWW}.
\par 
A few comments are in order regarding the relationship between
the calculations presented below and those presented in 
Ref.~\cite{yfsww2}. In Ref.~\cite{yfsww2}, we presented the
second order leading-log YFS exponentiated calculation
of the processes under discussion here and included for the
first time the complete YFS formfactor effect for the 
radiation from the $W^+W^-$ themselves, in a gauge 
invariant way~\cite{zep,argy}. With the advertised~\cite{LEP2YBWW}
precision goal of $0.5\%$ for the final LEP2 theoretical precision
tag on the signal prediction for the $W^+W^-$ pair production,
it is important to treat the complete ${\cal O}(\alpha)$ exactly
in the presence of our YFS exponentiation, including the exact
pure weak corrections at ${\cal O}(\alpha)$, as these may enter
at the level of this goal. It is for this reason that we are motivated to
carry-out our work in this paper.
\par
Specifically, what we will do is the following. In the next section,
we shall describe the extension the analysis in Ref.~\cite{yfsww2}
to include final state multiple photon radiation at the leading-log (LL)
level. This will lead to the development of the new YFS exponentiated
Monte Carlo event generator YFSWW3, involving as it will the 
extension of the YFS3 event generator in Ref.~\cite{yfs3}
for the process $e^+e^-\rightarrow f\bar f +n(\gamma)$ to
the process $e^+e^-\rightarrow W^+W^- +n(\gamma)$, where the
W-pair is then allowed to decay. This will set the stage
for our work in section 3, wherein we will include in the
YFSWW3 event generator the exact ${\cal O}(\alpha)$ results
for the hard photon residuals $\bar\beta_i,i=0,1$ in the
language of Ref.~\cite{yfs} (this language will be reviewed
briefly below) basing ourselves on the exact virtual electroweak
${\cal O}(\alpha)$ results and the 
exact ${\cal O}(\alpha)$ bremsstrahlung results of Refs.~\cite{jeger}. 
In section 4, we will then present some sample
Monte Carlo data to illustrate the size of the various 
levels of approximation that we access in our work, with
some focus on what is currently available in the literature
and how our results relate to it. Section 5 contains some concluding
remarks.
\par
\section{Multiple Initial + Final State Photon Radiative Effects}

In this section we discuss the relevant aspects of our YFS Monte Carlo
methods as they pertain to the problem of extending our YFS3
MC in Ref.~\cite{yfs3} for the processes 
$e^+e^-\rightarrow f\bar f + n(\gamma)$ in which both initial
and final state YFS exponentiated multiple photon radiation
is simulated to the $W^+W^-$ pair production processes of interest to us here.
We carry-out the respective extension as well to arrive at a
new MC event generator, YFSWW3, in which not only is the full
YFS formfactor effect with the $W$-pair contribution included 
as in YFSWW2~\cite{yfsww2} but
also real soft final state $n(\gamma)$ radiation from the W-pair itself
is calculated. Such results have not appeared elsewhere.
On the way we also set our notation and define our kinematics
for the work in the remaining sections.
\par
\begin{figure}[!ht]
\centering
\begin{picture}(48000,27000)
\thicklines
\bigphotons
\put(20000,12000){\oval(4000,9000)}
\put(18500,16000){\line(1,0){3000}}
\put(18200,15500){\line(1,0){3600}}
\multiput(18000,15000)(0,-500){13}{\line(1,0){4000}}
\put(18500, 8000){\line(1,0){3000}}
\put(18200, 8500){\line(1,0){3600}}
\THICKLINES
\drawline\fermion[\SE\REG](13500,20500)[6700]
\drawarrow[\NW\ATBASE](15000,19000)
\drawline\photon[\NE\FLIPPEDFLAT](\pmidx,\pmidy)[7]
\drawarrow[\E\ATBASE](18200,20500)
\put(11500,20000){\large $e^+$}
\put(13000,17500){\large $-p_1$}
\put(20500,22000){\large $\gamma_1$}
\put(17000,21500){\large $k_1$}
\drawline\fermion[\NE\REG](13500, 3500)[6700]
\drawarrow[\NE\ATTIP](15000, 5000)
\drawline\photon[\SE\FLAT](\pmidx,\pmidy)[7]
\drawarrow[\E\ATBASE](18200,3500)
\put(11500, 3000){\large $e^-$}
\put(14000, 6000){\large $q_1$}
\put(20500, 1500){\large $\gamma_n$}
\put(17000, 2000){\large $k_n$}
\drawline\photon[\E\REG](21500,16000)[7]
\drawline\fermion[\NE\REG](\photonbackx,\photonbacky)[3000]
\drawarrow[\NE\ATBASE](\pmidx,\pmidy)
\drawline\fermion[\SE\REG](\photonbackx,\photonbacky)[3000]
\drawarrow[\NW\ATBASE](\pmidx,\pmidy)
\put(26000,17000){\large $W^+$}
\put(26000,14500){\large $p_2$}
\put(31000,17500){$f_1$}
\put(31000,14000){$\bar{f_1'}$}
\put(28500,18500){ $p_{f_1}$}
\put(27500,13500){ $-p_{\bar{f_1'}}$}
\drawline\photon[\E\FLIPPED](21500, 8000)[7]
\drawline\fermion[\NE\REG](\photonbackx,\photonbacky)[3000]
\drawarrow[\SW\ATBASE](\pmidx,\pmidy)
\drawline\fermion[\SE\REG](\photonbackx,\photonbacky)[3000]
\drawarrow[\SE\ATBASE](\pmidx,\pmidy)
\put(26000, 9000){\large $W^-$}
\put(26000, 6500){\large $q_2$}
\put(31000, 9500){$\bar{f_2}$}
\put(31000, 6000){$f_2'$}
\put(27500,10500){ $-p_{\bar{f_2}}$}
\put(28500, 5500){ $p_{f_2'}$}
\multiput(20000,20500)(1000,-1000){5}{\circle*{350}}
\multiput(24200, 9000)(   0, 1500){5}{\circle*{350}}
\multiput(20000, 3500)(1000, 1000){5}{\circle*{350}}

\end{picture}

\caption{\small\sf 
The process 
\protect$e^+e^-\rightarrow W^+W^-\rightarrow 4 fermions = f_1+\bar{f'_1}
+\bar{f_2} + f_2'$, where 
\protect$\left({}^{f_i}_{f'_i}\right)_L$, \protect$i=1,2$, 
are SU$_{2L}$ doublets. Here, $p_A$ is the 4-momentum of $A$, 
\protect$A=f_i,f'_i,$
\protect$p_1(q_1)$ and $p_2(q_2)$ 
are the 4-momenta of \protect$e^+(e^-)$ and \protect$W^+(W^-)$
respectively. We use the notation \protect$C_L\equiv P_L\,C\equiv 
\frac{1}{2}(1-\gamma_5)\,C $ for any $C$.
}
\label{fig:WWprod}
\end{figure}
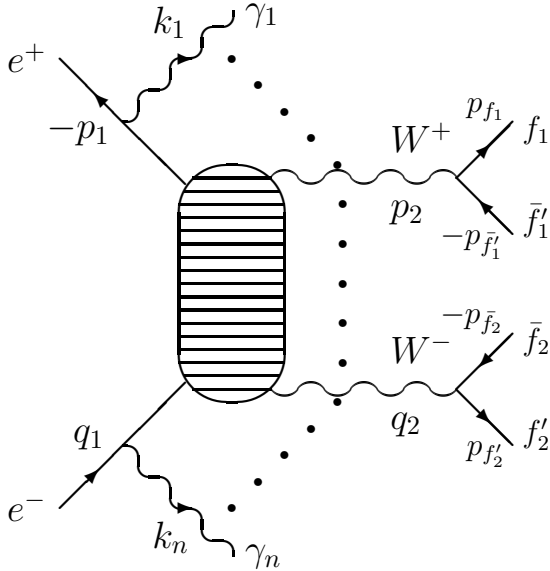

Specifically, the processes of interest to us here are illustrated
in Fig.~\ref{fig:WWprod}, where we have given our kinematics.

We consider $e^++e^-\rightarrow W^++W^-
+n(\gamma)\rightarrow 4 fermions + n(\gamma)$ at CMS energies $\sqrt{s}
\ge 2M_W$. In Ref.~\cite{yfsww2}, we have shown that
our YFS exponentiated Monte Carlo algorithms in Refs.~\cite{yfs2,yfs3} 
for the processes $e^+e^-\rightarrow f + \bar{f}+n(\gamma)$, where $f$ is a
fundamental fermion in the $SU_{2L}\times U_1$ theory~\cite{gsw}, has a gauge
invariant extension to $e^++e^-\rightarrow W^++W^-
+n(\gamma)\rightarrow 4 fermions + n(\gamma)$. 
In Ref.~\cite{yfsww2} we presented the extension for
our YFS2 initial state multiple photon Monte Carlo  
as the YFSWW2 multiple photon Monte Carlo in which
the full YFS form factor was taken into account. Here, we 
present the corresponding extension of our YFS3 initial + final
state multiple photon Monte Carlo in which mutiple photon
radiation from the $W$'s themselves is included in the 
in the lowest level of the Monte Carlo algorithm.
We refer to this extension of YFS3 to the $W$-pair production
process as YFSWW3 and it is available  from the authors \cite{yfsww3}.
\par
Specifically, on applying the results in Ref.~\cite{yfsww2}
for extending, to $W$-pair production in a gauge invariant way, 
our YFS formula for the 
process for $e^+e^-\rightarrow f + \bar{f}+n(\gamma)$
to our formula for the cross section in Ref.~\cite{yfs3},
we arrive at the gauge invariant YFS formula for the respective
$W$-pair production cross section in which multiple
initial state and multiple final state radiation 
is calculated, where we use the ratio of $\Gamma_W/M_W$ 
to justify treating the $W$-pair state as a ``final'' state on 
their mass shell
insofar as these radiative effects are concerned.
We then realize this formula by Monte Carlo methods
in complete analogy with what is done in Refs.~\cite{yfsww2,yfs3},
using as our Born cross section that given by
Ref.~\cite{hagi}. The resulting Monte Carlo, YFSWW3, 
at high energies compared to $M_W$ features final state
leading-log ${\cal O}(\alpha^2)$ radiative effects as well
as the initial state leading-log ${\cal O}(\alpha^2)$
radiative effects in YFSWW2. Thus, at NLC energies,
these final state radiative effects may be important
as we shall illustrate presently. In addition, the YFSWW3
final state $n(\gamma)$ radiation allows us to check the
technical precision of the $\bar\beta_0$-level $n(\gamma)$
soft radiation in YFSWW2 since the final state 
$n(\gamma)$ radiation is already present in the low level
background Monte Carlo in YFSWW3 whereas in YFSWW2 it is not.
We will also illustrate this presently.
\par

\let\sstl=\scriptscriptstyle
%
%
\def\Was{W\c as}
\def\Order#1{${\cal O}(#1$)}
\def\Ordpr#1{${\cal O}(#1)_{prag}$}
\def\bbe{\bar{\beta}}
\def\tbe{\tilde{\beta}}
\def\tal{\tilde{\alpha}}
\def\tom{\tilde{\omega}}
\def\half{ {1\over 2} }
\def\alf1{ {\alpha\over\pi} }

\def\Oaz{${\cal O}(\alpha^0)$}
\def\Oaf{${\cal O}(\alpha^1)$}
\def\Oas{${\cal O}(\alpha^2)$}


\begin{table}[!ht]

\centering
\begin{tabular}{||c|c|c|c|c||}
\hline\hline
 $E_{CM}\,[GeV]$  &  ISR & $+$ Coul.~corr. & $+\,Y'$-corr. &$+$ $WW$-rad.\\
\hline\hline
$175$ & $0.4906\pm 0.0002$ & $0.5046\pm 0.0002$ & $0.5053\pm 0.0002$ 
      & $0.5053\pm 0.0002$ \\
      & $0.4898\pm 0.0002$ & $0.5037\pm 0.0002$ & $0.5048\pm 0.0002$ 
      & $0.5048\pm 0.0002$ \\
\hline
$190$ & $0.6060\pm 0.0007$ & $0.6193\pm 0.0007$ & $0.6217\pm 0.0007$ 
      & $0.6219\pm 0.0009$ \\
      & $0.6034\pm 0.0007$ & $0.6166\pm 0.0007$ & $0.6195\pm 0.0007$
      & $0.6197\pm 0.0009$ \\
\hline
$205$ & $0.6359\pm 0.0008$ & $0.6480\pm 0.0008$ & $0.6514\pm 0.0008$ 
      & $0.6516\pm 0.0010$ \\
      & $0.6315\pm 0.0008$ & $0.6436\pm 0.0008$ & $0.6476\pm 0.0008$ 
      & $0.6475\pm 0.0010$ \\
\hline\hline
$500$ & $0.2910\pm 0.0003$ & $0.2946\pm 0.0003$ & $0.2970\pm 0.0004$ 
      & $0.2954\pm 0.0004$ \\
      & $0.3538\pm 0.0004$ & $0.3582\pm 0.0004$ & $0.3591\pm 0.0004$ 
      & $0.3571\pm 0.0005$ \\
\hline\hline
\end{tabular}

\caption{\small\sf The results of the $10^5$ (except for $E_{CM}=175\,GeV$,
where it is $10^6$) statistics sample (unweighted events)
from YFSWW3 for the total cross section $\sigma\,[pb]$. 
The upper results at each value of energy are for the Standard Model
couplings constants, while the lower ones are for anomalous couplings
constants (\protect$\delta\kappa=\delta\lambda=0.1$). 
See the text for more details.
}
\label{tab:xsec}
\end{table}

More precisely, in Table~\ref{tab:xsec},
we show the result of our YFSWW3.0 simulation of the process
in Fig.~\ref{fig:WWprod} with the $c\bar{s} + e\bar{\nu}_e$
final state, for CMS energies 175 GeV, 190 GeV, 205 GeV and
500 GeV which have LEP2 and NLC in mind.
We present results according to the notation in 
Ref.~\cite{yfsww2} so that ISR denotes initial state $n(\gamma)$
radiation, ``Coul. corr.'' denotes that the Coulomb correction
after the fashion of Ref.~\cite{Coul} is included,
``$Y'$-corr.'' denotes that the full YFS form-factor effect already
featured in Ref.~\cite{yfsww2} is turned on,
and finally ``$WW$-rad.'' denotes the new feature
of YFSWW3.0 in which the real soft 
$n(\gamma)$ bremsstrahlung from the 
$W$-pair is included in the simulation (already at the level
of the low level Monte Carlo algorithm in the sense of
Ref.~\cite{yfs2} for example).
In addition, the upper and lower results
in each entry in the table correspond to the case
of Standard Model and anomalous ($\delta\kappa=\delta\lambda=0.1$)
$WWV$ couplings in the notation of Ref.~\cite{hagi}.
Thus, the comparison of the last two columns shows that
the effects of the final state $n(\gamma)$, while negligible
at LEP2 energies, is significant at NLC energies.
Comparison of the first three columns in the table
with the analogous results in the respective Table~1
of Ref.~\cite{yfsww2} shows that indeed we have a very good
agreement between YFSWW2 and YFSWW3.0 in the calculation
of effects in which the real final state radiation should
not be important. This represents a very good technical precision
check on the two calculations. Moreover, it shows
that YFSWW3.0 is an excellent starting point for 
developing the exact ${\cal O}(\alpha)$ YFS exponentiated
$\bar\beta_1$-level Monte Carlo event generator calculation
in which one has an exact  ${\cal O}(\alpha)$ calculation
in the presence of initial $+$ final state ${\cal O}(\alpha^2)$
leading-log radiation with YFS exponentiation -- such a calculation
has not appeared elsewhere. To this we now turn in the
next section.
\par

\section{ Exact ${\cal O}(\alpha)$ YFS Exponentiated
WW-Pair Production: YFSWW3.1}

In this section, we develop the exact ${\cal O}(\alpha)$ realization
of the hard photon residuals $\bar\beta_n, n=0,1$ in our YFSWW3
Monte Carlo event generator, where we refer to 
Refs.~\cite{yfs,yfs2,yfsww2} for a precise definition of these 
residuals. We start with $\bar\beta_0$.\par
For the construction of the exact ${\cal O}(\alpha)$ electroweak result for
$\bar\beta_0$, which we denote by $\bar\beta_0^{(1)}$,
we first note that
\begin{equation}
{1\over 2}\bar\beta_0^{(1)} = d\sigma^{one-loop}/d\Omega 
                         -2\alpha \Re B d\sigma_B/d\Omega
\label{eq1}
\end{equation}
where the YFS virtual infrared function $B$ for the process
$e^++e^-\rightarrow W^++W^-$ is defined in Ref.~\cite{yfsww2},
where the cross section $d\sigma_B/d\Omega$ is the respective
Born cross section, and where $d\sigma^{one-loop}/d\Omega$ is the respective
exact one-loop correction to the cross section given by the results of 
Ref.~\cite{jeger}. We have implemented the formula in (\ref{eq1})
in our YFSWW3 Monte Carlo event generator to arrive at a realization of
the exact ${\cal O}(\alpha)$ electroweak result for the hard photon
residual $\bar\beta_0$. 
\par
Turning now to the exact ${\cal O}(\alpha)$ result for the hard
photon residual $\bar\beta_1$, which we denote by
$\bar\beta_1^{(1)}$, we first note that
\begin{equation}
{1\over 2}\bar\beta_1^{(1)} = d\sigma^{B1}/kdkd\Omega_{\gamma}d\Omega
                              -\tilde S(k)d\sigma_B/d\Omega
\label{eq2}
\end{equation}
where $\tilde S(k)$ is the respective YFS real infrared function
defined in Ref.~\cite{yfsww2} and $d\sigma^{B1}/kdkd\Omega_{\gamma}d\Omega$
is the respective exact ${\cal O}(\alpha)$ bremsstrahlung cross
section\footnote{We would like to thank K. Ko\l{}odziej for providing us with
                 routines for the ${\cal O}(\alpha)$ hard bremsstrahlung 
                 matrix element.} 
from Ref.~\cite{jeger}
for the photon into the phase space element $kdkd\Omega_{\gamma}$
when the $W^-$ is produced into the solid angle $d\Omega$, for example.
We have implemented the result (\ref{eq2}) into our YFSWW3 Monte Carlo
as well and the resulting version of it, including both
the results in (\ref{eq1}) and in (\ref{eq2}) is version 3.1: YFSWW3.1.
In the next section, we illustrate some of its applications.
\par

\section{Illustrative Results from YFSWW3.1}

In this section we illustrate the application of the exact
${\cal O}(\alpha)$ YFS exponentiated Monte Carlo event generator
YFSWW3.1 to the WW-pair production and decay at LEP2 and NLC
energies. We continue to use the $c\bar{s} + e\bar{\nu}_{e}$
4-fermion final state for definiteness.
\par
Specifically, we have recorded in Table~\ref{tab:approx} a summary of the
sizes of the various approximations which we have realized 
in YFSWW3.1 including the exact ${\cal O}(\alpha)$ YFS exponentiated
result.
\begin{table}[!ht]

\centering
\begin{tabular}{||c|c|c|c|c|c||}
\hline\hline
 $E_{CM}\,[GeV]$  & $\sigma_0\,[pb] $ 
& $(\sigma_2^{prag} - \sigma_0)/\sigma_0$  
& $(\sigma_2^{LL} - \sigma_1^{LL})/\sigma_0$
& $(\sigma_1^{ex} - \sigma_1^{LL})/\sigma_0$
& $(\sigma_1^{ex} - \sigma_1^{ap})/\sigma_0$ 
\\
\hline\hline
$161$ & $0.1768$ & $-24.6\%$ & $+0.11\%$ & $-0.80\%$ & $+0.21\%$ \\
$175$ & $0.5891$ & $-15.4\%$ & $+0.12\%$ & $-1.31\%$ & $-0.007\%$ \\
$190$ & $0.6792$ & $-10.0\%$ & $+0.11\%$ & $-1.70\%$ & $-0.22\%$ \\
$205$ & $0.6850$ & $ -7.1\%$ & $+0.10\%$ & $-2.20\%$ & $-0.61\%$ \\
\hline
$500$ & $0.2710$ & $ +4.4\%$ & $+0.19\%$ & $-4.65\%$ & $-3.08\%$ \\
      &          & $ +4.4\%$ & $+0.19\%$ & $-5.05\%$ & $-3.08\%$ \\
\hline\hline
\end{tabular}

\caption{\small\sf Various contributions (approximations) to the 
YFS exponentiated $WW$ cross section as a fraction [in \%] 
of the Born cross section $\sigma_0$: 
$\sigma_2^{prag}$ denotes the so-called pragmatic 
\protect${\cal O}(\alpha^2)$ cross section (see the text for more details),
$\sigma_1^{LL}$ and $\sigma_2^{LL}$ are the LL approximations of
\protect${\cal O}(\alpha^1)$ and \protect${\cal O}(\alpha^2)$, respectively,
$\sigma_1^{ex}$ is the exact \protect${\cal O}(\alpha^1)$ result,
while $\sigma_1^{ap}$ is the so-called Improved Born Approximations
(see the text for more details). 
The results in the lower line for \protect$E_{CM}=500\,GeV$ correspond to 
the situation when LL QED corrections for the $W^+W^-$ state 
are also included in the corresponding LL contributions. 
}
\label{tab:approx}
\end{table}
 
We show results for CMS energies $E_{CM}$ of 161, 175, 190, 205
and 500~GeV. The definitions of the 
YFS exponentiated cross sections in the table are
as follows: $\sigma_2^{prag}$ denotes that
the cross section contains the exact ${\cal O}(\alpha)$
correction and the ${\cal O}(\alpha^2)$ initial state 
LL correction at the level of the
$W$-pair production process; $\sigma_n^{LL}$ denotes that the
cross section contains initial state LL QED radiative corrections through 
${\cal O}(\alpha^n)$;
$\sigma_n^{ex}$ denotes that the cross section contains the 
exact radiative correction through ${\cal O}(\alpha^n)$; 
$\sigma_n^{ap}$ indicates that the cross section
contains an approximate treatment of the radiative corrections through
${\cal O}(\alpha^n)$ (so-called Improved Born Approximation)
and we only discuss the case $n=1$ where we 
have implemented the approximate cross section of Ref.~\cite{sigap}.
The second entry for $E_{CM} = 500$ GeV corresponds in each column to
a different treatment of the LL QED radiation in which both initial
and final state LL QED effects are included -- it gives us an estimate
of the size of the sub-leading QED radiative effects at the higher
orders when it is compared with the first entry for example.  
The cross sections shown in the table correspond to simulations of
$10^6$ (weighted) events at all energies except for the 
$E_{CM} = 500$ GeV results, where $10^7$ events were simulated.  
What we see in the table is that
the ${\cal O}(\alpha^2)$ LL effects are well below the $0.5\%$ targeted
precision tag of the Physics at LEP2 ``WW Cross-Sections and Distributions''
Physics Working Group~\cite{LEP2YBWW} for all LEP2 energies. For the 
500 GeV case, the high precision objectives of certain NLC physics
issues would clearly necessitate inclusion of these second order
LL effects. The difference between the exact ${\cal O}(\alpha)$
result $\sigma_1^{ex}$ and the first order LL result $\sigma_1^{LL}$
shows that even at LEP2 energies, the non-leading QED and 
pure EW corrections are important~\cite{sigap}; and, as advertised
in Ref.~\cite{sigap}, the use of the approximation
$\sigma_1^{ap}$ at LEP2 energies would give results within the 
desired $0.5\%$ precision tag except perhaps at the highest LEP2 CMS energy
of 205 GeV, where the difference between the exact result and the
$\sigma_1^{ap}$ is $0.61\%$, just beyond the desired tag. 
Thus, for all but the highest LEP2 energies, the $\sigma_1^{ap}$ could
be used to realize a gain in CPU time for event generation of a
factor of $\sim 30$ without an unacceptable loss in precision
relative a $0.5\%$ precision tag. At the
NLC type energy $E_{CM} = 500$ GeV the approximation $\sigma_1^{ap}$
is not supposed to be accurate and our last column bears this out,
although we see that it is better than the $\sigma_1^{LL}$ at such
high energies. We are encouraged that our best result, $\sigma_2^{prag}$,
is not very sensitive to the inclusion of the final state LL QED
${\cal}(\alpha^2)$ effects at 500 GeV; this is not true of $\sigma_1^{LL}$,
as expected. We stress that, in the final state, $L=ln (s/M_W^2)-1$ is
only 2.67 for $E_{CM} = 500$ GeV so that the LL series
in the final state is one of several effects of comparable significance
in comparison to the LL initial state series at ${\cal}(\alpha^2)$
and we show the last row in the table to illustrate the size of
such effects. \par
 
\begin{table}[!ht]

\centering
\begin{tabular}{||c|c|c|c|c|c|c|c||}
\hline\hline
 $E_{CM}\,[GeV]$  
& $(\bbe_0^{(1)} - \bbe_{0,LL}^{(1)})/\sigma_0$ 
& $(\bbe_1^{(1)} - \bbe_{1,LL}^{(1)})/\sigma_0 $ \\
\hline\hline
$161$ & $-0.85\%$ & $+0.04\%$ \\
$175$ & $-1.24\%$ & $-0.07\%$ \\
$190$ & $-1.57\%$ & $-0.13\%$ \\
$205$ & $-2.06\%$ & $-0.15\%$ \\
\hline
$500$ & $-5.38\%$ & $+0.73\%$ \\
      & $-6.05\%$ & $+1.00\%$ \\
\hline\hline
\end{tabular}

\caption{\small\sf Differences between exact and LL approximated non-infrared
contributions to the YFS exponentiated $WW$ cross section as a fraction
[in \%] of the Born cross section $\sigma_0$.
The results in the lower line for \protect$E_{CM}=500\,GeV$ correspond to 
the situation when LL QED corrections for the $W^+W^-$ state 
are also included in the corresponding LL contributions. 
}
\label{tab:bety}
\end{table}

Since we have incorporated the exact ${\cal O}(\alpha)$ correction
in our calculation, it is instructive to look at the size of its
effect in comparison to the LL QED radiative correction insofar as
our YFS hard photon residuals are concerned. We show this in
Table~\ref{tab:bety}. In the table, the notation is such that
$\bbe_n^{(m)}$ is the ${\cal O}(\alpha^m)$ result for the hard YFS
photon residual $\bbe_n$ and the subscript $LL$ denotes that
it is computed in the LL approximation (in the initial state LL approximation
to be precise). What we see in the table is that the main effect
of the exact result in comparison to the LL one 
is realized in the hard photon residual $\bbe_0$; the effect on
$\bbe_1$ in comparison to the LL approximation is well below the 
$0.5\%$ precision tag of Ref.~\cite{LEP2YBWW} for LEP2 energies.
For the NLC energy 500 GeV, we see that, while the main difference between
the exact result
and the LL approximation is still in $\bbe_0$, that in $\bbe_1$
is also significant for high precision NLC physics objectives.
\par 
\begin{figure}[!ht]
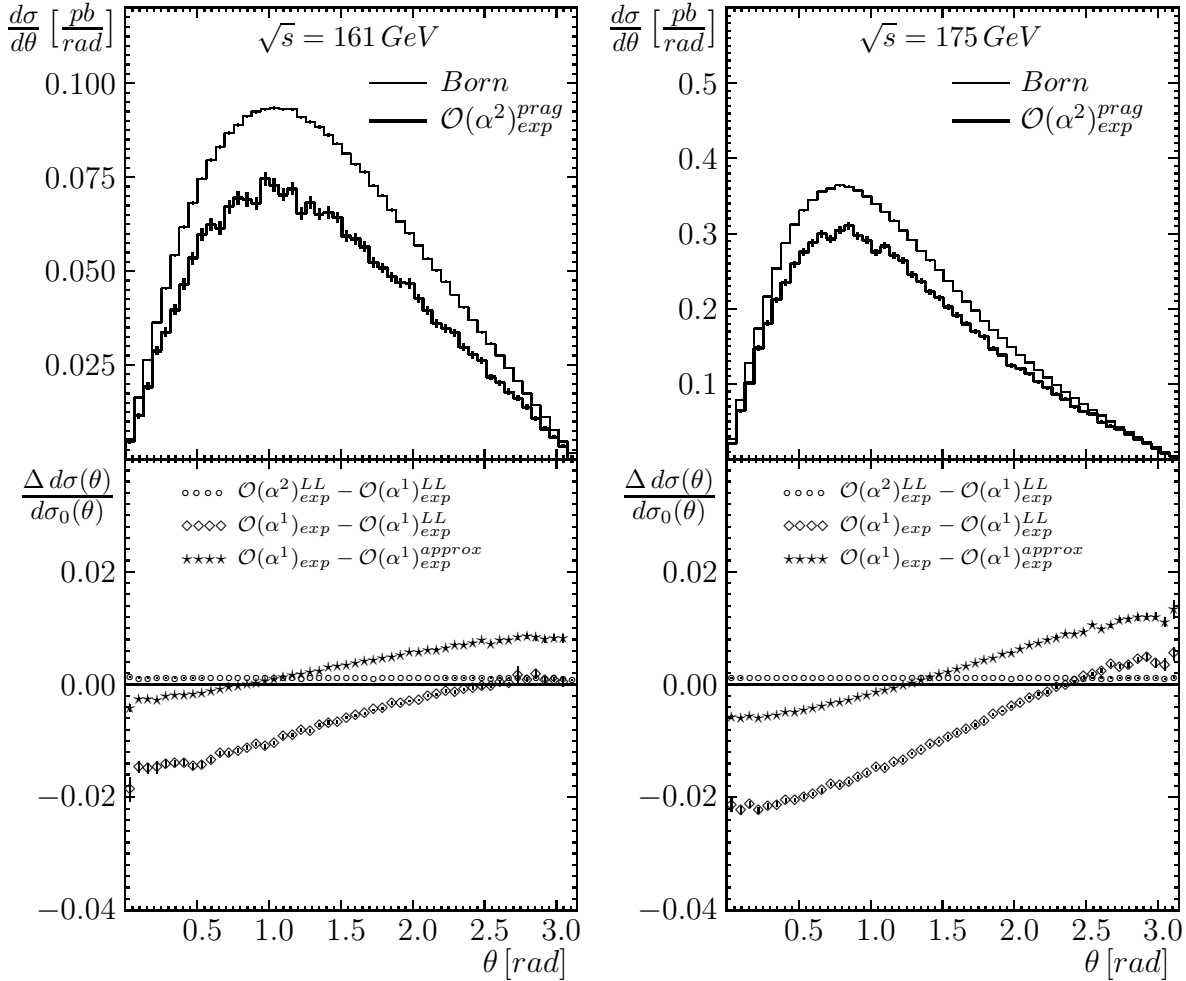

\centering

\setlength{\unitlength}{0.05mm}
} 

\end{picture} 

\caption[fig1]{\small\sf 
The $W^-$ angular distributions for two LEP2 energies: 
\protect$\sqrt{s}=161$ GeV (left picture) and 
\protect$\sqrt{s}=175$ GeV (right picture).
The upper parts of the pictures show the differential cross
sections in Born (thin-lines) and 
in \protect${\cal O}(\alpha^2)_{exp}^{prag}$ (thick-lines)
approximations, where the latter one denotes 
the YFS exponentiated cross section including the exact 
\protect${\cal O}(\alpha^1)$ matrix element together
with the 2-nd order LL ISR terms (see the text for more details).
In the lower part of each picture the following contributions 
to the YFS exponentiated cross section (divided by the corresponding 
Born differential cross section \protect$d\sigma_0$) are presented: 
the 2-nd order LL ISR contribution (\protect$\circ\circ$), 
the 1-st order non-LL contribution including electroweak
radiative corrections (\protect$\diamond\diamond$) and 
the difference between the exact \protect${\cal O}(\alpha^1)$ 
matrix elements and the so-called Improved Born Approximation 
(\protect$\star\star$). 
The last one shows the quality of IBA for the differential cross section
as a function of the $W$-boson polar angle.   
}
\label{fig:LEP2a}

\end{figure} 

\begin{figure}[!ht]
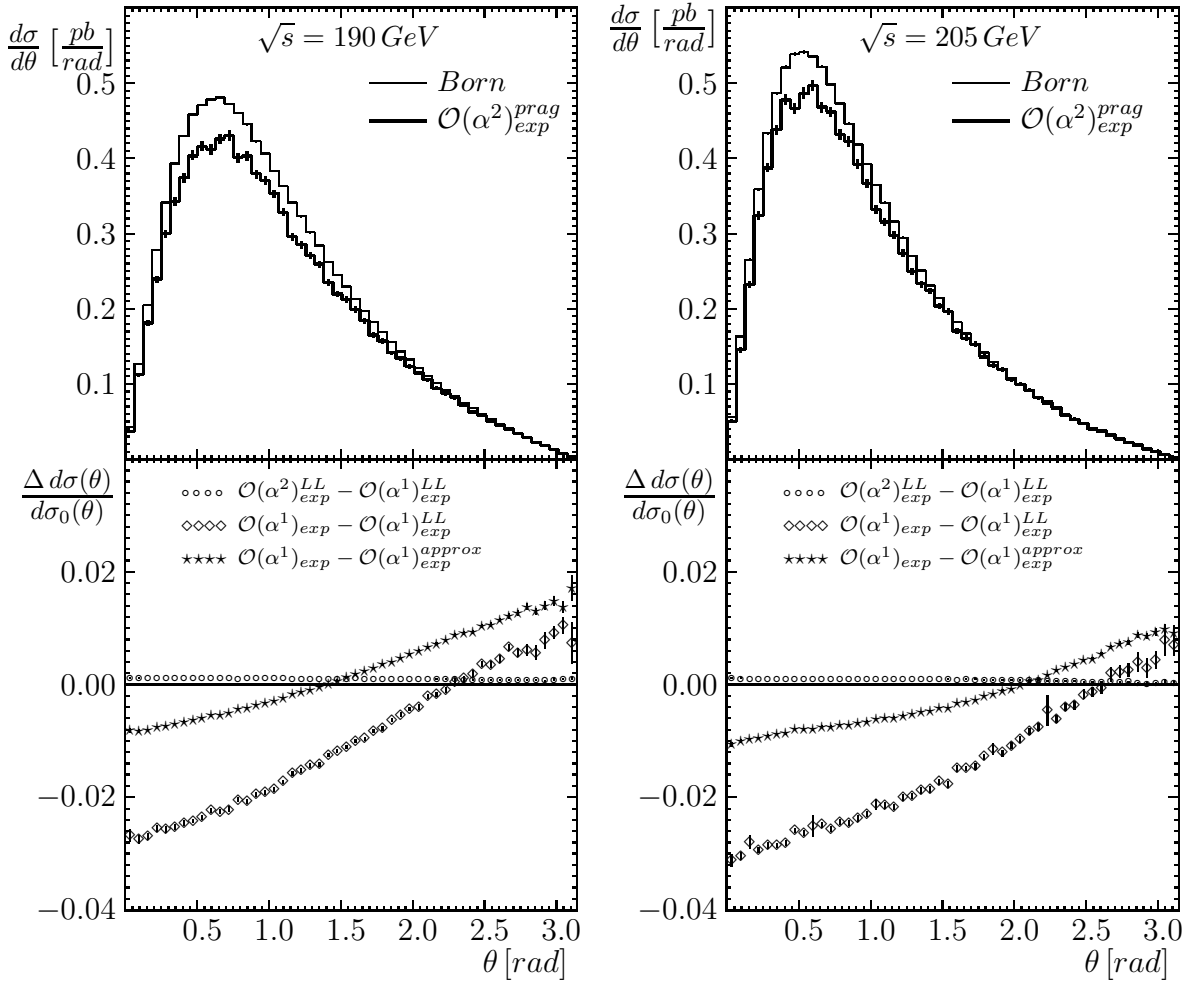

\centering
\setlength{\unitlength}{0.05mm}
} 

\end{picture} 

\caption[fig2]{\small\sf 
The same as in Fig.~\ref{fig:LEP2a} but for other LEP2 energies: 
\protect$\sqrt{s}=190$ GeV (left picture) and 
\protect$\sqrt{s}=205$ GeV (right picture).
}
\label{fig:LEP2b}
\end{figure} 

\begin{figure}[!ht]
\centering
\setlength{\unitlength}{0.05mm}
\begin{picture}(3200,1500)
\put(300,250){\begin{picture}( 1200,1200)
\put(0,0){\framebox( 1200,1200){ }}
\multiput(  190.99,0)(  190.99,0){   6}{\line(0,1){25}}
\multiput(     .00,0)(   19.10,0){  63}{\line(0,1){10}}
\multiput(  190.99,1200)(  190.99,0){   6}{\line(0,-1){25}}
\multiput(     .00,1200)(   19.10,0){  63}{\line(0,-1){10}}
\put( 191,-25){\makebox(0,0)[t]{ $    0.5 $}}
\put( 382,-25){\makebox(0,0)[t]{ $    1.0 $}}
\put( 573,-25){\makebox(0,0)[t]{ $    1.5 $}}
\put( 764,-25){\makebox(0,0)[t]{ $    2.0 $}}
\put( 955,-25){\makebox(0,0)[t]{ $    2.5 $}}
\put(1146,-25){\makebox(0,0)[t]{ $    3.0 $}}
\put(1050,-100){\makebox(0,0)[t]{ $\theta\,[rad] $}}
\multiput(0,     .00)(0,  300.00){   5}{\line(1,0){25}}
\multiput(0,   30.00)(0,   30.00){  40}{\line(1,0){10}}
\multiput(1200,     .00)(0,  300.00){   5}{\line(-1,0){25}}
\multiput(1200,   30.00)(0,   30.00){  40}{\line(-1,0){10}}
\put(-25,   0){\makebox(0,0)[r]{ $    0.0 $}}
\put(-25, 300){\makebox(0,0)[r]{ $    0.2 $}}
\put(-25, 600){\makebox(0,0)[r]{ $    0.4 $}}
\put(-25, 900){\makebox(0,0)[r]{ $    0.6 $}}
\put(-25,1150){\makebox(0,0)[r]{\large 
               $\frac{d\sigma}{d\theta}\,[\frac{pb}{rad}] $}}
\end{picture}}

\put(650,1300){\small $\sqrt{s}=500\,GeV$} 
\put(700, 950){\begin{picture}( 600,300)
\put(0  ,200){\line(1,0){150} }
\thicklines
\put(0  ,100){\line(1,0){150} }
\put(180,180){\small $Born$ }
\put(180, 80){\small ${\cal O}(\alpha^2)_{exp}^{prag}$ }
\thinlines
\end{picture}}
\put(300,250){\begin{picture}( 1200,1200)
\thinlines 
\newcommand{\x}[3]{\put(#1,#2){\line(1,0){#3}}}
\newcommand{\y}[3]{\put(#1,#2){\line(0,1){#3}}}
\newcommand{\z}[3]{\put(#1,#2){\line(0,-1){#3}}}
\newcommand{\e}[3]{\put(#1,#2){\line(0,1){#3}}}
\y{   0}{   0}{ 688}\x{   0}{ 688}{  48}
\e{  24}{  688}{   0}
\y{  48}{ 688}{  44}\x{  48}{ 732}{  48}
\e{  72}{  732}{   0}
\z{  96}{ 732}{ 264}\x{  96}{ 468}{  48}
\e{ 120}{  468}{   0}
\z{ 144}{ 468}{ 145}\x{ 144}{ 323}{  48}
\e{ 168}{  323}{   0}
\z{ 192}{ 323}{  89}\x{ 192}{ 234}{  48}
\e{ 216}{  233}{   0}
\z{ 240}{ 234}{  60}\x{ 240}{ 174}{  48}
\e{ 264}{  174}{   0}
\z{ 288}{ 174}{  43}\x{ 288}{ 131}{  48}
\e{ 312}{  131}{   0}
\z{ 336}{ 131}{  31}\x{ 336}{ 100}{  48}
\e{ 360}{  100}{   0}
\z{ 384}{ 100}{  22}\x{ 384}{  78}{  48}
\e{ 408}{   78}{   0}
\z{ 432}{  78}{  17}\x{ 432}{  61}{  48}
\e{ 456}{   61}{   0}
\z{ 480}{  61}{  12}\x{ 480}{  49}{  48}
\e{ 504}{   49}{   0}
\z{ 528}{  49}{   9}\x{ 528}{  40}{  48}
\e{ 552}{   40}{   0}
\z{ 576}{  40}{   7}\x{ 576}{  33}{  48}
\e{ 600}{   33}{   0}
\z{ 624}{  33}{   6}\x{ 624}{  27}{  48}
\e{ 648}{   27}{   0}
\z{ 672}{  27}{   4}\x{ 672}{  23}{  48}
\e{ 696}{   23}{   0}
\z{ 720}{  23}{   4}\x{ 720}{  19}{  48}
\e{ 744}{   19}{   0}
\z{ 768}{  19}{   4}\x{ 768}{  15}{  48}
\e{ 792}{   15}{   0}
\z{ 816}{  15}{   3}\x{ 816}{  12}{  48}
\e{ 840}{   12}{   0}
\z{ 864}{  12}{   3}\x{ 864}{   9}{  48}
\e{ 888}{    9}{   0}
\z{ 912}{   9}{   2}\x{ 912}{   7}{  48}
\e{ 936}{    7}{   0}
\z{ 960}{   7}{   2}\x{ 960}{   5}{  48}
\e{ 984}{    5}{   0}
\z{1008}{   5}{   2}\x{1008}{   3}{  48}
\e{1032}{    3}{   0}
\z{1056}{   3}{   1}\x{1056}{   2}{  48}
\e{1080}{    2}{   0}
\z{1104}{   2}{   1}\x{1104}{   1}{  48}
\e{1128}{    1}{   0}
\z{1152}{   1}{   1}\x{1152}{   0}{  48}
\e{1176}{    0}{   0}
\end{picture}} 
\put(300,250){\begin{picture}( 1200,1200)
\thicklines 
\newcommand{\x}[3]{\put(#1,#2){\line(1,0){#3}}}
\newcommand{\y}[3]{\put(#1,#2){\line(0,1){#3}}}
\newcommand{\z}[3]{\put(#1,#2){\line(0,-1){#3}}}
\newcommand{\e}[3]{\put(#1,#2){\line(0,1){#3}}}
\y{   0}{   0}{ 668}\x{   0}{ 668}{  48}
\e{  24}{  667}{   4}
\y{  48}{ 668}{  79}\x{  48}{ 747}{  48}
\e{  72}{  745}{   4}
\z{  96}{ 747}{ 255}\x{  96}{ 492}{  48}
\e{ 120}{  490}{   4}
\z{ 144}{ 492}{ 153}\x{ 144}{ 339}{  48}
\e{ 168}{  338}{   2}
\z{ 192}{ 339}{  94}\x{ 192}{ 245}{  48}
\e{ 216}{  244}{   2}
\z{ 240}{ 245}{  64}\x{ 240}{ 181}{  48}
\e{ 264}{  180}{   2}
\z{ 288}{ 181}{  43}\x{ 288}{ 138}{  48}
\e{ 312}{  137}{   2}
\z{ 336}{ 138}{  32}\x{ 336}{ 106}{  48}
\e{ 360}{  105}{   2}
\z{ 384}{ 106}{  24}\x{ 384}{  82}{  48}
\e{ 408}{   82}{   2}
\z{ 432}{  82}{  15}\x{ 432}{  67}{  48}
\e{ 456}{   66}{   0}
\z{ 480}{  67}{  13}\x{ 480}{  54}{  48}
\e{ 504}{   54}{   0}
\z{ 528}{  54}{   9}\x{ 528}{  45}{  48}
\e{ 552}{   45}{   0}
\z{ 576}{  45}{   7}\x{ 576}{  38}{  48}
\e{ 600}{   37}{   0}
\z{ 624}{  38}{   5}\x{ 624}{  33}{  48}
\e{ 648}{   32}{   0}
\z{ 672}{  33}{   5}\x{ 672}{  28}{  48}
\e{ 696}{   27}{   0}
\z{ 720}{  28}{   5}\x{ 720}{  23}{  48}
\e{ 744}{   23}{   0}
\z{ 768}{  23}{   3}\x{ 768}{  20}{  48}
\e{ 792}{   19}{   0}
\z{ 816}{  20}{   3}\x{ 816}{  17}{  48}
\e{ 840}{   17}{   0}
\z{ 864}{  17}{   2}\x{ 864}{  15}{  48}
\e{ 888}{   14}{   0}
\z{ 912}{  15}{   3}\x{ 912}{  12}{  48}
\e{ 936}{   11}{   0}
\z{ 960}{  12}{   3}\x{ 960}{   9}{  48}
\e{ 984}{    9}{   0}
\z{1008}{   9}{   1}\x{1008}{   8}{  48}
\e{1032}{    8}{   0}
\z{1056}{   8}{   2}\x{1056}{   6}{  48}
\e{1080}{    6}{   0}
\z{1104}{   6}{   2}\x{1104}{   4}{  48}
\e{1128}{    4}{   0}
\z{1152}{   4}{   3}\x{1152}{   1}{  48}
\e{1176}{    1}{   0}
\end{picture}} 

\put(1900,250){\begin{picture}( 1200,1200)
\put(0,0){\framebox( 1200,1200){ }}
\multiput(  190.99,0)(  190.99,0){   6}{\line(0,1){25}}
\multiput(     .00,0)(   19.10,0){  63}{\line(0,1){10}}
\multiput(  190.99,1200)(  190.99,0){   6}{\line(0,-1){25}}
\multiput(     .00,1200)(   19.10,0){  63}{\line(0,-1){10}}
\put( 191,-25){\makebox(0,0)[t]{ $    0.5 $}}
\put( 382,-25){\makebox(0,0)[t]{ $    1.0 $}}
\put( 573,-25){\makebox(0,0)[t]{ $    1.5 $}}
\put( 764,-25){\makebox(0,0)[t]{ $    2.0 $}}
\put( 955,-25){\makebox(0,0)[t]{ $    2.5 $}}
\put(1146,-25){\makebox(0,0)[t]{ $    3.0 $}}
\put(1050,-100){\makebox(0,0)[t]{ $\theta\,[rad] $}}
\multiput(0,     .00)(0,  200.00){   7}{\line(1,0){25}}
\multiput(0,   20.00)(0,   20.00){  60}{\line(1,0){10}}
\multiput(1200,     .00)(0,  200.00){   7}{\line(-1,0){25}}
\multiput(1200,   20.00)(0,   20.00){  60}{\line(-1,0){10}}
\put(0,  600.00){\line(1,0){1200}}
\put(-25,   0){\makebox(0,0)[r]{ $   -0.15 $}}
\put(-25, 200){\makebox(0,0)[r]{ $   -0.10 $}}
\put(-25, 400){\makebox(0,0)[r]{ $   -0.05 $}}
\put(-25, 600){\makebox(0,0)[r]{ $    0.00 $}}
\put(-25, 800){\makebox(0,0)[r]{ $    0.05 $}}
\put(-25,1000){\makebox(0,0)[r]{ $    0.10 $}}
\put(-25,1150){\makebox(0,0)[r]{\large 
               $\frac{\Delta\,d\sigma(\theta)}{d\sigma_{0}(\theta)} $}}
\end{picture}}

\put(2050,1100){\begin{picture}( 600,300)
\multiput(10 ,210)(30,0){4}{\circle{15} }
\multiput(0  ,110)(30,0){4}{\scriptsize $\diamond$ }
\multiput(0  , 20)(30,0){4}{\scriptsize $\star$ }
\put(150,200){\scriptsize
${\cal O}(\alpha^2)_{exp}^{LL} - {\cal O}(\alpha^1)_{exp}^{LL}$ }
\put(150,110){\scriptsize
${\cal O}(\alpha^1)_{exp} - {\cal O}(\alpha^1)_{exp}^{LL}$ }
\put(150, 20){\scriptsize
${\cal O}(\alpha^1)_{exp} - {\cal O}(\alpha^1)_{exp}^{approx}$ }
\end{picture}}
\put(1900,250){\begin{picture}( 1200,1200)
\newcommand{\R}[2]{\put(#1,#2){\circle{ 15}}}
\newcommand{\Ee}[3]{\put(#1,#2){\line(0,1){#3}}}
\R{  24}{ 605}
\Ee{  24}{  605}{   0}
\R{  72}{ 605}
\Ee{  72}{  605}{   0}
\R{ 120}{ 606}
\Ee{ 120}{  606}{   0}
\R{ 168}{ 608}
\Ee{ 168}{  607}{   0}
\R{ 216}{ 608}
\Ee{ 216}{  608}{   0}
\R{ 264}{ 609}
\Ee{ 264}{  608}{   0}
\R{ 312}{ 609}
\Ee{ 312}{  609}{   0}
\R{ 360}{ 610}
\Ee{ 360}{  609}{   0}
\R{ 408}{ 610}
\Ee{ 408}{  610}{   0}
\R{ 456}{ 611}
\Ee{ 456}{  611}{   0}
\R{ 504}{ 612}
\Ee{ 504}{  612}{   0}
\R{ 552}{ 613}
\Ee{ 552}{  612}{   0}
\R{ 600}{ 613}
\Ee{ 600}{  613}{   0}
\R{ 648}{ 615}
\Ee{ 648}{  615}{   2}
\R{ 696}{ 615}
\Ee{ 696}{  615}{   2}
\R{ 744}{ 617}
\Ee{ 744}{  616}{   2}
\R{ 792}{ 620}
\Ee{ 792}{  619}{   2}
\R{ 840}{ 627}
\Ee{ 840}{  626}{   2}
\R{ 888}{ 632}
\Ee{ 888}{  631}{   2}
\R{ 936}{ 643}
\Ee{ 936}{  642}{   2}
\R{ 984}{ 664}
\Ee{ 984}{  662}{   4}
\R{1032}{ 693}
\Ee{1032}{  690}{   6}
\R{1080}{ 745}
\Ee{1080}{  741}{   8}
\R{1128}{ 802}
\Ee{1128}{  795}{  14}
\R{1176}{ 848}
\Ee{1176}{  836}{  24}
\end{picture}} 
\put(1900,250){\begin{picture}( 1200,1200)
\newcommand{\R}[2]{\put(#1,#2){\makebox(0,0){\scriptsize $\diamond$}}}
\newcommand{\Ee}[3]{\put(#1,#2){\line(0,1){#3}}}
\R{  24}{ 485}
\Ee{  24}{  484}{   4}
\R{  72}{ 455}
\Ee{  72}{  454}{   4}
\R{ 120}{ 420}
\Ee{ 120}{  417}{   6}
\R{ 168}{ 385}
\Ee{ 168}{  382}{   4}
\R{ 216}{ 369}
\Ee{ 216}{  364}{  10}
\R{ 264}{ 344}
\Ee{ 264}{  338}{  12}
\R{ 312}{ 322}
\Ee{ 312}{  317}{  10}
\R{ 360}{ 316}
\Ee{ 360}{  309}{  14}
\R{ 408}{ 322}
\Ee{ 408}{  313}{  16}
\R{ 456}{ 308}
\Ee{ 456}{  300}{  18}
\R{ 504}{ 332}
\Ee{ 504}{  321}{  22}
\R{ 552}{ 344}
\Ee{ 552}{  327}{  34}
\R{ 600}{ 356}
\Ee{ 600}{  341}{  30}
\R{ 648}{ 388}
\Ee{ 648}{  366}{  44}
\R{ 696}{ 386}
\Ee{ 696}{  360}{  54}
\R{ 744}{ 351}
\Ee{ 744}{  333}{  36}
\R{ 792}{ 418}
\Ee{ 792}{  376}{  84}
\R{ 840}{ 409}
\Ee{ 840}{  370}{  80}
\R{ 888}{ 438}
\Ee{ 888}{  376}{ 124}
\R{ 936}{ 313}
\Ee{ 936}{  293}{  40}
\R{ 984}{ 338}
\Ee{ 984}{  277}{ 122}
\R{1032}{ 212}
\Ee{1032}{  160}{ 104}
\R{1176}{  58}
\Ee{1176}{ -313}{ 742}
\end{picture}} 
\put(1900,250){\begin{picture}( 1200,1200)
\newcommand{\R}[2]{\put(#1,#2){\makebox(0,0){\scriptsize $\star$}}}
\newcommand{\Ee}[3]{\put(#1,#2){\line(0,1){#3}}}
\R{  24}{ 571}
\Ee{  24}{  571}{   0}
\R{  72}{ 552}
\Ee{  72}{  552}{   0}
\R{ 120}{ 516}
\Ee{ 120}{  516}{   0}
\R{ 168}{ 475}
\Ee{ 168}{  474}{   0}
\R{ 216}{ 433}
\Ee{ 216}{  432}{   2}
\R{ 264}{ 393}
\Ee{ 264}{  392}{   2}
\R{ 312}{ 355}
\Ee{ 312}{  354}{   2}
\R{ 360}{ 326}
\Ee{ 360}{  324}{   2}
\R{ 408}{ 302}
\Ee{ 408}{  301}{   4}
\R{ 456}{ 280}
\Ee{ 456}{  278}{   4}
\R{ 504}{ 266}
\Ee{ 504}{  264}{   4}
\R{ 552}{ 251}
\Ee{ 552}{  249}{   4}
\R{ 600}{ 249}
\Ee{ 600}{  247}{   4}
\R{ 648}{ 241}
\Ee{ 648}{  239}{   4}
\R{ 696}{ 239}
\Ee{ 696}{  237}{   4}
\R{ 744}{ 230}
\Ee{ 744}{  228}{   4}
\R{ 792}{ 228}
\Ee{ 792}{  225}{   6}
\R{ 840}{ 221}
\Ee{ 840}{  218}{   6}
\R{ 888}{ 212}
\Ee{ 888}{  209}{   6}
\R{ 936}{ 210}
\Ee{ 936}{  206}{   8}
\R{ 984}{ 220}
\Ee{ 984}{  216}{   8}
\R{1032}{ 259}
\Ee{1032}{  254}{  10}
\R{1080}{ 309}
\Ee{1080}{  302}{  14}
\R{1128}{ 422}
\Ee{1128}{  412}{  18}
\R{1176}{ 532}
\Ee{1176}{  524}{  16}
\end{picture}} 

\end{picture} 

\caption[fig3]{\small\sf 
The same as in Fig.~\ref{fig:LEP2a} but for the NLC energy: 
\protect$\sqrt{s}=500$ GeV. 
}
\label{fig:NLC}

\end{figure}
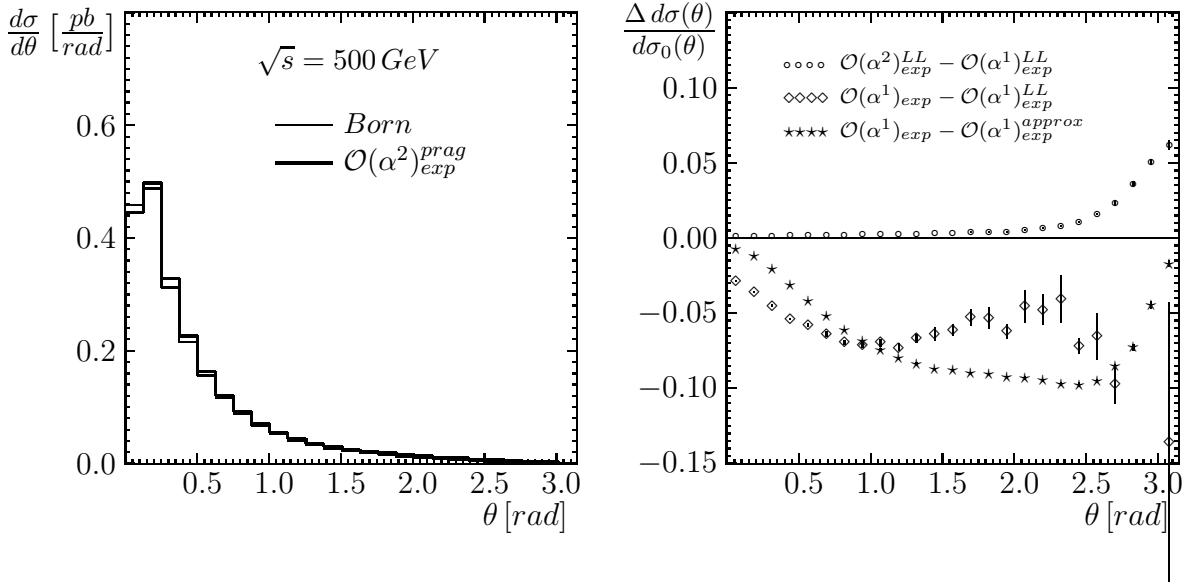 
These results on the total cross sections are then analyzed in more
detail in Figs.~\ref{fig:LEP2a}, \ref{fig:LEP2b}, \ref{fig:NLC}, wherein we
show the $W^-$ angular distributions corresponding to the 
respective $E_{CM}$ values in Tables~1-3. 
For each value of $E_{CM}$ we show the $W^-$ angular distribution 
at the Born level and at the ${\cal O}(\alpha^2)^{prag}_{exp}$ level 
($\sigma^{prag}_2$ level) as well
as the differences $\Delta d\sigma(\theta)_{2-1}=
d\sigma^{LL}_2(\theta)-d\sigma^{LL}_1(\theta)$,
$\Delta d\sigma(\theta)_{ex-1}=d\sigma^{ex}_1(\theta)-d\sigma^{LL}_1(\theta)$,
and 
$\Delta d\sigma(\theta)_{ex-ap}=d\sigma^{ex}_1(\theta)-d\sigma^{ap}_1(\theta)$
in ratio to
the respective Born level differential cross section $d\sigma_0(\theta)$,
where Fig.~1 contains this information for $E_{CM}= 161, 175$ GeV,
Fig.~2 contains it for $E_{CM}= 190, 205$ GeV, and Fig.~3 contains
it for $E_{CM}= 500$ GeV. In the figures, we denote
by ${\cal O}(\alpha^n)^{LL}_{exp}$ the YFS exponentiated cross section 
$\sigma^{LL}_n$ level, by ${\cal O}(\alpha^n)_{exp}$ 
the YFS exponentiated cross section $\sigma^{ex}_n$ level,
and by  ${\cal O}(\alpha^n)^{approx}_{exp}$
the YFS exponentiated cross section $\sigma^{ap}_n$ level.
The differential cross section results are fully consistent
with the total cross section results shown above. Moreover, they
show us that at all energies, the  radiative corrections are
significant throughout the entire angular distribution. At LEP2
energies, the angular dependence of the ratio $\Delta d\sigma(\theta)_{2-1}/
d\sigma_0(\theta)$ is essentially flat; at the NLC energy
500 GeV, it is nontrival and strongly varying in the backward direction.
The ratios $\Delta d\sigma(\theta)_{ex-1}/d\sigma_0(\theta)$ and
$\Delta d\sigma(\theta)_{ex-ap}/d\sigma_0(\theta)$ are smoothly
varying in $\theta$ at LEP2 and at NLC energies; at all energies,
in the region where the cross sections are largest, the latter is
smaller in magnitude than the former, as expected; away from this
region, especially near the backward direction, the opposite is true.
Of course, at the NLC energy 500 GeV, one does not expect 
the cross section $\sigma^{ap}_1$ be as accurate as it is
at the LEP2 energies for which it was optimized 
and this is borne-out both in the
$\theta$ dependence of the ratio
$\Delta d\sigma(\theta)_{ex-ap}/d\sigma_0(\theta)$
as well as in the total cross section results in Table 2.
Not shown in the Figs.~2-4 are the corresponding results for
the initial + final state LL YFS exponentiated approximation
at 500~ GeV; its curves are entirely similar to those 
shown in Fig.~4 so that we do not present them separately here.\par
Our conlusion is that the exact results show that for
the targeted precision of the Physics at LEP2 Workshop
the LL approximation is inadequate, both for the total cross section
and for the differential cross section
; the non-leading corrections
must be taken into account. In YFSWW3.1, we have the only
multiple photon YFS exponentiated amplitude based calculation
of these effects in which the respective infrared singularities are
canceled to all orders in $\alpha$ and in which the
corrections themselves are realized on an event-by-event basis.
\par
\section{ Conclusions}

In this paper we have presented and illustrated
the first ever 4-fermion Monte Carlo event generator
for $W^+W^-$ pair production and decay
in which the full EW ${\cal O}(\alpha)$ correction is
calculated in the presence 
of amplitude based YFS exponentiation in which infrared
singularities are canceled to all orders in $\alpha$.
The resulting Monte Carlo event generator, YFSWW3.1,
is available from the authors \cite{yfsww3}.
\par
As we have illustrated by showing the results of simulations
at several LEP2 energies and at the NLC energy 500 GeV, 
the nonleading ${\cal O}(\alpha)$ EW corrections are indeed important and
must be taken into account, both for the overall normalization
and for the detailed angular distributions. At 
LEP2 energies, we showed that the approximate 
representation in Ref.~\cite{sigap} of these non-leading
corrections does indeed stay with $0.61$ of the 
exact ${\cal O}(\alpha)$ EW result in the presence
of our YFS exponentiation as we realize it by Monte Carlo
methods; for the NLC energy, this is no longer the case, as
expected. We stress that these last remarks apply both to
the total cross sections and to the differential cross sections
in the regions where they have their largest values. 
Our $\bbe_1$ level second order LL YFS exponentiated
results are thus a significant improvement over
the $\bbe_0$ level results in Ref.~\cite{yfsww2}; it is also
a significant improvement
over the other available Monte Carlo
event generators~\cite{dimaWW,ohl} in the literature; for, none of them
feature a 4-fermion, amplitude based, exact EW ${\cal}O(\alpha)$
YFS exponentiated calculation in which infrared singularities
are canceled to all orders in $\alpha$, on an event-by-event basis
as we do in YFSWW3.1. We look forward with excitement to
the many applications of our work.\par

\vspace{7mm}
\noindent 
{\large\bf  Acknowledgements}

Two of us (S. J. and B.F.L. W.) acknowledge the
kind hospitality of Prof. G. Veneziano and the CERN Theory 
Division while this work was completed. One of us
(B.F.L.W.) acknowledges the support of Prof. D. Schlatter
and the ALEPH Collaboration
in the final stages of this work.


 

\begin{thebibliography}{99}
%
\bibitem{gsw}
S.L. Glashow, Nucl. Phys. {\bf 22} (1961) 579; \\
S. Weinberg, Phys. Rev. Lett. {\bf 19} (1967) 1264; \\
A. Salam, in {\em Elementary Particle Theory}, ed. N. Svartholm
(Almqvist and Wiksells, Stockholm, 1968), p 367.
%
\bibitem{LEP2YBWW}
W. Beenakker et al.,
{\em WW Cross-Sections and Distributions}, 
in Reports of CERN Workshop ``Physics at LEP2'',
edited by G. Altarelli, T. Sj\"ostrand, and F. Zwirner 
(CERN, Geneva, 1996), Yellow Report CERN 96-01, Vol.~1, p.~79.
%
\bibitem{yfs}
D. R. Yennie, S. Frautschi and H. Suura, Ann. Phys. {\bf 13} (1961) 379.
%
\bibitem{yfsww2}
S. Jadach, W. P{\l}aczek, M. Skrzypek and B.F.L. Ward,
Phys. Rev. {\bf D54} (1996) 5434.
%
\bibitem{zep}
U. Baur and D. Zeppenfeld, Phys. Rev. Lett. {\bf 75} (1995) 1002.
%
\bibitem{argy}
E.N. Argyres {\it et al.}, Phys. Lett. {\bf B358} (1995) 339.
%
\bibitem{yfs3}
S. Jadach and B.F.L. Ward, Phys. Lett. {\bf B274} (1992) 470;
{\em YFS3 Monte Carlo Program Long Write-Up}, in preparation
(YFS3 is available from the authors).
%
\bibitem{jeger}
J. Fleischer,  F. Jegerlehner and M. Zra\l{}ek, 
Z. Phys. {\bf C42} (1989) 409;\\
M. Zra\l{}ek and K. Ko\l{}odziej, Phys. Rev. {\bf D43} (1991) 43; \\
J. Fleischer, K. Ko\l{}odziej and F. Jegerlehner, 
Phys. Rev. {\bf D47} (1993) 830; \\
J. Fleischer {\it et al.}, Comp. Phys. Commun. {\bf 85} (1995) 29; 
and references therein.
%
\bibitem{yfs2}
S. Jadach and B.F.L. Ward,
Comput. Phys. Commun. {\bf 56} (1990) 351.
%
\bibitem{yfsww3}
S. Jadach, W. P{\l}aczek, M. Skrzypek and B.F.L. Ward,
{\em The MC Event Generator YFSWW3},
available at the WWW URL http://enigma.phys.utk.edu/pub/YFSWW/.
%
\bibitem{hagi}
K. Hagiwara, R. D. Peccei, D. Zeppenfeld and   K. Hikasa,
Nucl. Phys. {\bf B282} (1987) 253.
%
\bibitem{Coul}
V.S. Fadin, V.A. Khoze, A.D. Martin and W.J. Stirling,
Phys. Lett. {\bf B363} (1995) 112, and references therein.
%
\bibitem{sigap}
S. Dittmaier, M. B\"{o}hm 
and A. Denner, Nucl. Phys. {\bf B376} (1992) 29;
E: {\bf B391} (1993) 483.
%
\bibitem{dimaWW}
D. Bardin {\it et al.},``Event Generators for WW Physics'', in
\underline{ Physics at LEP2},
edited by G. Altarelli, T. Sj\"ostrand, and F. Zwirner 
(CERN, Geneva, 1996), Yellow Report CERN 96-01, Vol.~2, p.~3.
%
\bibitem{ohl}
T. Ohl, preprint IKDA-96-21, 1996, to appear
in \underline{Proc. 1996 Cracow Int. Symp.}\newline 
\underline{ Rad. Corr.}, ed. S. Jadach,
(Institute of Physics of 
the Polish Academy of Sciences,
Cracow, 1997), in press.
%
%
%
%
%
%
%
%
%
%
%
%
%
%
%
%
%
%
%
%
%
\end{thebibliography}
\end{document}